\documentclass[12pt]{iopart}

\usepackage{array}
\usepackage{graphicx}
\usepackage[
backend=biber,
sorting=none
]{biblatex}
\begin{document}

\title[Depth resolved study of annealing in IrMn/(Fe, Co, CoFe) exchange bias systems]{Depth resolved study of annealing in IrMn/(Fe, Co, CoFe) exchange bias systems}

\author{Yu. Khaydukov$^1$, A. Dobrynin$^1$, S. Hassan$^1$, M. Ormston$^1$, K. Nikolaev$^2$, P. Bencok$^3$, R. Fan$^3$, P. Steadman$^3$, A. Csik$^4$, A. Vorobiev$^5$}

\address{$^1$Seagate Technology, Derry, Northern Ireland, BT48 0BF, UK}
\address{$^2$Seagate Technology, Bloomington, MN 55435, USA}
\address{$^3$Diamond Light Source, Didcot, Oxfordshire, OX11 0DE, UK}
\address{$^4$HUN-REN Institute for Nuclear Research (ATOMKI), Debrecen,  Hungary}
\address{$^5$Department of Physics and Astronomy, Uppsala University, Uppsala, Sweden}

\begin{abstract}
Depth resolved study of structural and magnetic profiles of antiferromagnetic/ferromagnetic (AFM/FM) system upon annealing was performed in this work. We studied systems comprising of AFM IrMn and FM (Co, Fe, Co$_{70}$Fe$_{30}$) structures using polarized neutron and soft X-ray scattering, secondary neutral spectrometry, and magnetometry. Structural depth profiles obtained from neutron reflectometry indicate non-homogeneity of the AFM layer even before annealing, which is associated with the migration of manganese to the surface of the sample. Annealing of samples with CoFe and Co layers leads to a slight increase ($\sim$ 5 \%) in the migration of manganese, which, however, does not lead to significant degradation of the exchange coupling at the AFM/FM interface. A significantly different picture was observed in the Fe/IrMn systems where a strong migration of iron into the AFM layer was observed upon annealing of the sample, leading to erosion of the magnetic profile, the formation of a non-magnetic alloy and degradation of the pinning strength. This study can be useful in the design of AF/FM systems in different spintronics devices, including HDD read heads, where thermal annealing is applied at different stages of the device manufacturing process.
\end{abstract}

\section{Introduction}

Exchange biased (EB) antiferromagnetic (AFM) – ferromagnetic (FM) bilayers are essential parts of modern spintronic devices, such as hard disk drive (HDD) read sensors or magnetic random access memory (MRAM) \cite{nogues99,radu08}. Exchange coupling at the AFM/FM interface is the key factor determining the pinning strength of the FM layer in single crystalline systems. In polycrystalline EB systems used in real applications, such as IrMn alloys, granular structure of the AFM is of high importance. Setting the EB direction in industrially relevant systems is performed by AFM/FM layers’ deposition and/or subsequent annealing in applied magnetic field. Typically, a few annealing steps at different temperatures and magnetic fields are required during the device fabrication. Such annealing may alter both the interface structure and AFM grain alignment, thus affecting the pinning strength \cite{OGrady2010} . Other works report on significant pinning strength variation in single crystalline FM/AFM bilayers depending on the IrMn AFM structure, which in turn depends on deposition and annealing conditions \cite{Kohn2013}. Thus, a detailed depth-resolved study of the processes occurring at the AFM/FM interface during annealing is an important task for both fundamental science and applied research.

Depth-resolved techniques such as Rutherford backscattering (RBS), Auger, or mass spectrometry are traditionally employed to analyze element-specific depth profiles near the AFM/FM interface. For instance, the application of Auger and RBS spectrometry in the examination of IrMn/CoFe(NiFe) systems has revealed substantial manganese migration to the surface upon annealing, attributed to manganese’s high oxygen affinity \cite{Lee01,Lee02}.

Polarized neutron reflectometry (PNR) is a powerful method for studying magnetic heterostructures, which makes it possible to obtain magnetization depth profiles in addition to chemical profiles \cite{Daillant2008}. The experimentally measured reflectivities of neutrons with different polarizations make it possible to reconstruct depth profiles of  nuclear and magnetic scattering length densities (SLDs). PNR has been extensively used to study the magnetic state of AFM/FM systems (see e.g. \cite{Khaydukov21} and references therein) and the structural/magnetic alterations occurring during annealing and/or irradiation \cite{Lund2004,Kirby2007,Causer2018,Quarterman2019}. In the last cited work of Quarterman et al. \cite{Quarterman2019} a detailed examination of MnN/CoFeB systems is presented. The authors show that the diffusion of nitrogen to the underlying Ta seed layer during annealing leads to both structural and magnetic changes near the AFM/FM system. The use of a complementary X-ray Magnetic Circular Dichroism (XMCD) method made it possible to obtain additional elemental sensitivity to atoms of the AFM and FM layers. 

In this study, we employed PNR and complementary methods (XMCD, magnetometry, and mass spectrometry) to conduct a detailed examination of the structural and magnetic changes of such widely used in industry materials  as antiferromagnetic IrMn$_3$, ferromagnetic Co, Fe and their alloy.

\section{Samples preparation and characterization techniques}

\begin{figure}[ht]
\centering
\includegraphics[width=0.9\textwidth]{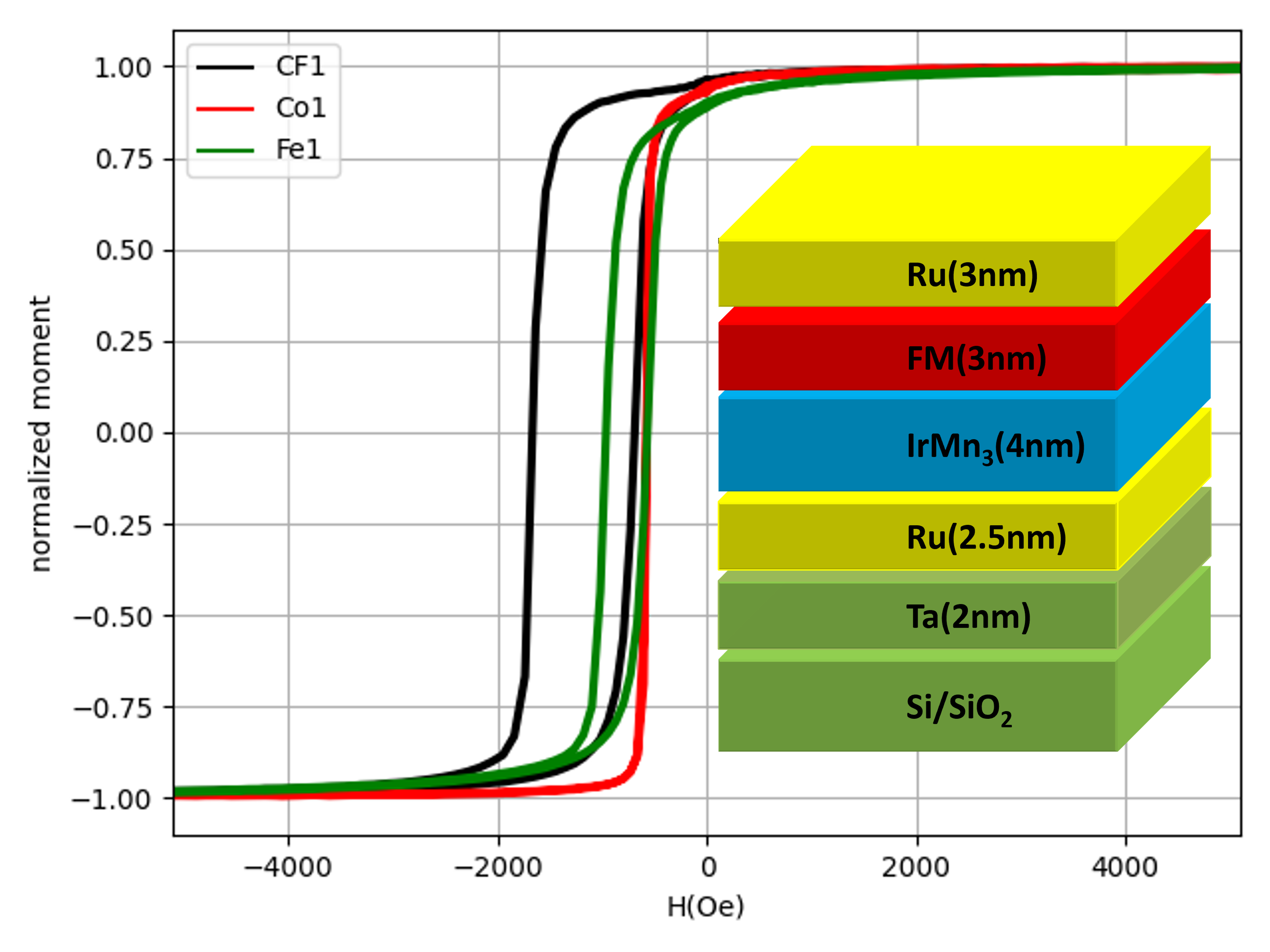}
\caption{Normalized hysteresis loops for chosen set of samples. Inset shows sketch of the samples used in study.}
\label{fig1}
\end{figure}

For the scattering experiments we have prepared series of samples on thermally oxidized silicon substrates with nominal composition Ta(2nm)/Ru(2.5nm)/IrMn$_3$(4nm) FM(3nm)/Ru(3nm). Sketch of the samples is shown in inset to Fig.\ref{fig1}, details of deposition can be can be found elsewhere \cite{ODonnell2019}. Here polycrystalline disordered gamma IrMn$_3$ (IM) is an antiferromagnetic layer used to pin ferromagnetic one. For the latter we used pure Co, Fe and Co$_{70}$Fe$_{30}$ alloy. To pin the FM layer, part of the samples were annealed for 6 hours at the applied field of 0.18T at temperatures 250°C and 300°C in UHV conditions (see table \ref{tab:magnetic_properties}). The choice of annealing temperatures was dictated by the typical thermal stability region for these systems, ensuring absence of strong intermixing at the interfaces \cite{Lee01,Nikolayev2016,Quarterman2019}.
Also, for SNMS we have prepared sample with slightly thicker CoFe(4nm) and IM(6nm) layers as the method requires certain volume for confident detection.
PNR experiments have been performed at the monochromatic reflectometer Super ADAM at ILL \cite{Devishvili2013,Vorobiev2015} at room temperature using polarizer-only set-up (polarization 99.8\%, wavelength 5.21$\AA$). In-plane magnetic field of $H$=6kOe was applied during experiment normal to the scattering surface.
To extract nuclear and magnetic depth profile experimental PNR data were fit to the model. Fit procedure consists of minimization of Figure of Merit: $FOM=1/Nlog_{10}(R_{exp}-R_{th})^2$, where $R_{th}$ are theoretical reflectivities calculated for a model depth profile, constructed based on nuclear and magnetic SLD, thicknesses of the layers and rms roughness at every interface. For the visual representation of the fit results it is convenient to use so-called spin asymmetry (SA): $S = (R_+ - R_-)/(R_+ + R_-)$.

Soft X-ray magnetic circular dichroism (XMCD) measurements were performed at the I10 beamline, Diamond Light Source. The measurements were performed at normal beam incidence in a high field magnet end station at room temperature at zero field and at 2T field, applied along the beam propagation. This field was sufficient to saturate the samples. Total electron yield (TEY) detection was used to obtain the absorption spectra. As TEY is only sensitive to first few nm from the sample surface, and the IrMn layer we were only able to obtain reliable XMCD signals from Co and Fe, but not from Mn L2/3 edges. 

Magnetization measurements were performed using a ‘looper’ extraction magnetometer, allowing to measure the whole 200mm wafer, as well as a Quantum Design PPMS VSM magnetometer. Fig.\ref{fig1} shows typical exchange biased hysteresis loops. The values of exchange bias field and exchange coupling energy density were extracted from such hysteresis loop for all samples (Table \ref{tab:magnetic_properties}). As can be seen from the table, the interfacial exchange interaction (IEC) energy density for samples with Co and CoFe layers decreases marginally (by 0.02 $erg/cm^2$) after annealing at 300°C, as compared to 250°C, similar to previous observation [14]. Fe/IM systems in contrast shows a significant degradation of exchange coupling upon annealing, with IEC reduced by 30\% for the sample annealed at 300°C as compared to that annealed at 250°C.

\begin{table}[ht]
\centering
\caption{Coercivity, exchange bias field, and IEC energy density of the samples.}
\label{tab:magnetic_properties}
\begin{tabular}{| l | l | l | l | l | l | }
\hline
Sample ID & FM & Annealed & Hc(kOe) & Heb (kOe) & IEC ($erg/cm^2$) \\
\hline
CF0 & CoFe(3nm) & No & 0.435 & - & - \\
CF1 & CoFe(3nm) & 250°C & 0.405 & -1.452 & 0.74 \\
CF2 & CoFe(3nm) & 300°C & 0.578 & -1.371 & 0.72 \\
Co0 & Co(3nm) & No & 0.134 & - & - \\
Co1 & Co(3nm) & 250°C & 0.078 & -0.636 & 0.29 \\
Co2 & Co(3nm) & 300°C & 0.078 & -0.645 & 0.27 \\
Fe0 & Fe(3nm) & No & 0.346 & - & - \\
Fe1 & Fe(3nm) & 250°C & 0.289 & -0.922 & 0.42 \\
Fe2 & Fe(3nm) & 300°C & 0.219 & -0.8 & 0.29 \\
\hline
\end{tabular}
\end{table}

\section{PNR data analysis}
\subsection{Samples with CoFe}
a) Treatment of the as-deposited sample CF0\\
Prior to analyzing the neutron data, let’s recall that neutron reflectometry is sensitive to the contrast of nuclear SLDs of adjacent layers. The nuclear SLD of a layer is the product of the average packing density of atoms and the averaged scattering length of all atoms comprising the layer. Table \ref{tab:SLs} shows the scattering lengths of the elements presented in our samples. 
\begin{table}[ht]
\centering
\caption{Coherent scattering lengths of atoms.}
\label{tab:SLs}
\begin{tabular}{|c|c|c|c|c|c|c|c|c|c|}
\hline
    Element & Ru & Co & Fe & Ir & Mn &  Ta & Si & O \\ \hline
    scattering length (fm) & 7.03 & 2.49 & 9.45 &  10.6 & -3.73 & 6.91 & 4.15 & 5.803 \\ \hline
%    SLD (10$^{-4}$nm$^{-2}$) & 5.20 & - & - &  - & - & 3.82 & - & - \\ \hline
  \end{tabular}
 \end{table}

From the table one can make several quantitative conclusions. First of all, due to negative scattering length of Mn the average scattering length of the IrMn$_3$ layer is quite close (with an accuracy of about 5\%) to zero. Second, The scattering lengths of Ru and Ta are quite close to each other, therefore, in the neutron experiment, these layers will be indistinguishable. The same holds for Fe and Ir, hence the interdiffusion of these elements at the AFM/FM interface will not change SLDs of the corresponding layers (although will definitely change the magnetization of the FM layer)

\begin{figure}[ht]
\centering
\includegraphics[width=0.9\textwidth]{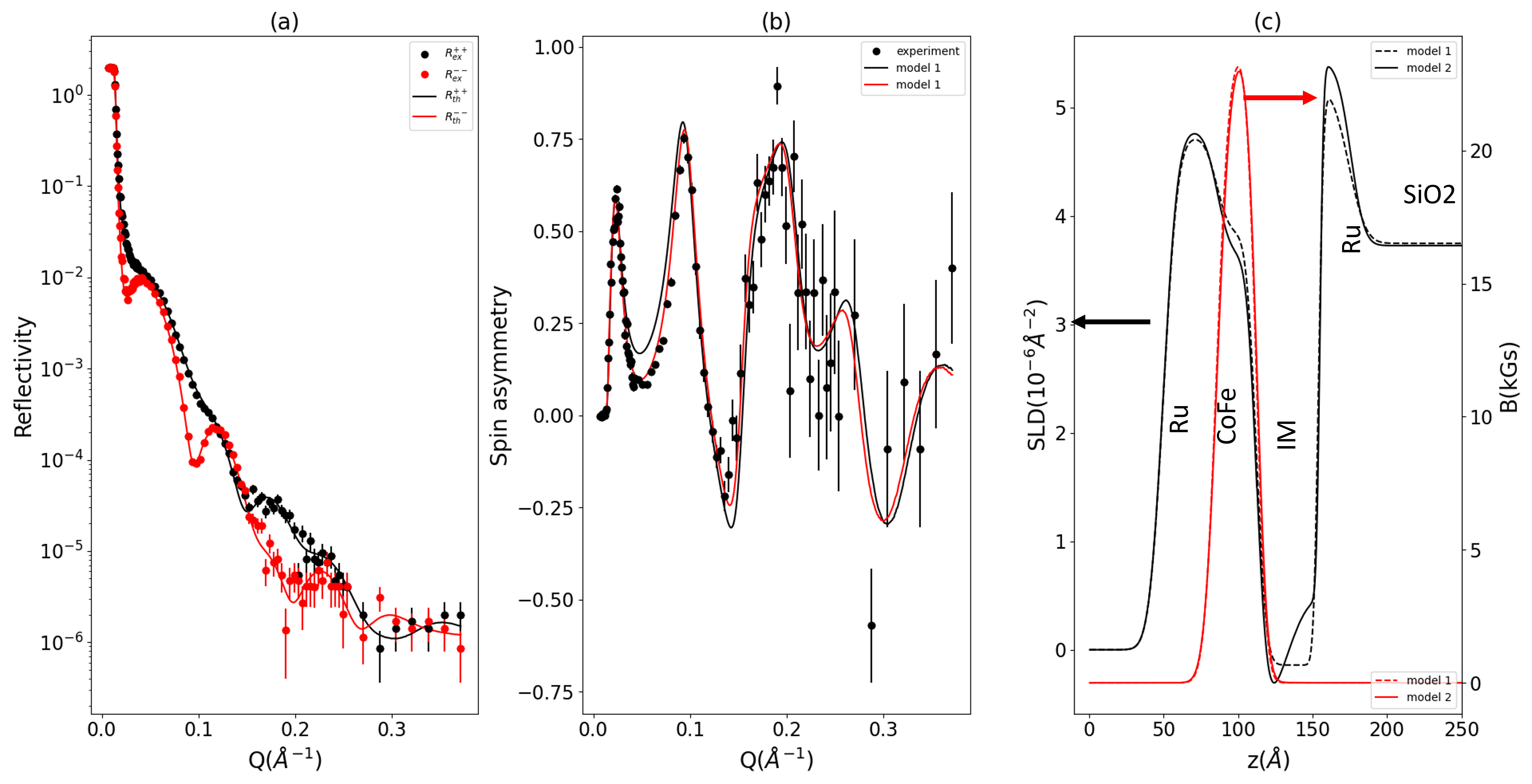}
\caption{(a) Experimental (dots) and model (solid lines) reflectivities for sample CF0. (b) Experimental (dots) and model curves (solid lines) for two models. Nuclear (black lines) and magnetic (red lines) depth profiles of these models are shown in (c).}
\label{fig2}
\end{figure}

We start analysis with the as-deposited sample CF0. The reflectivity curves are shown in Fig.\ref{fig2}a. In our initial model (model 1) we varied parameters of the nuclear potential (thickness, SLD and roughness) and the magnetization of the CoFe layer. The best-fit spin asymmetry with $FOM=1.82 \cdot 10^{-2}$ is shown in  Fig.\ref{fig2}b with black line. As one can see, the model curve qualitatively describe experiment, although there are quite big discrepancies around second oscillation with maximum at $Q \approx 0.1 \AA ^{-1}$. Nuclear and magnetic depth profiles for this model are shown with black and red dashed lines in Fig.\ref{fig1}c. As it follows from the figure, magnetization in the centre of CoFe layer is within the accuracy of fitting close to the bulk value of 22kG (or 2.2T in SI units). Looking ahead, we can say that all measured samples in as-deposited state have magnetisation in center of the FM layers close to corresponding bulk values. We have obtained typical rms roughness of $5-7 \AA$ which, however, due to rather small thickness of the layers leads to observable smearing of the depth profiles. In more advanced model 2 we split the AFM layer onto two part and varied SLD of both parts independently. This allowed us to improve $FOM=1.55 \cdot 10^{-2}$ (see red line in Fig.\ref{fig2}b) and better describe experimental data around the second oscillation. Comparison of nuclear depth profiles for model 1 and 2 (Fig. 1c) shows that division of the IM layer onto two sublayers resulted in an inhomogeneous profile having negative value $-1.3 \cdot 10^{-6} \AA ^{-2}$ near the interface with CoFe layer and gradually increasing to $2.0 \cdot 10^{-6} \AA ^{-2}$ near interface with bottom Ru layer.  \\

b) Effect of annealing.\\

\begin{figure}[ht]
\centering
\includegraphics[width=0.8\textwidth]{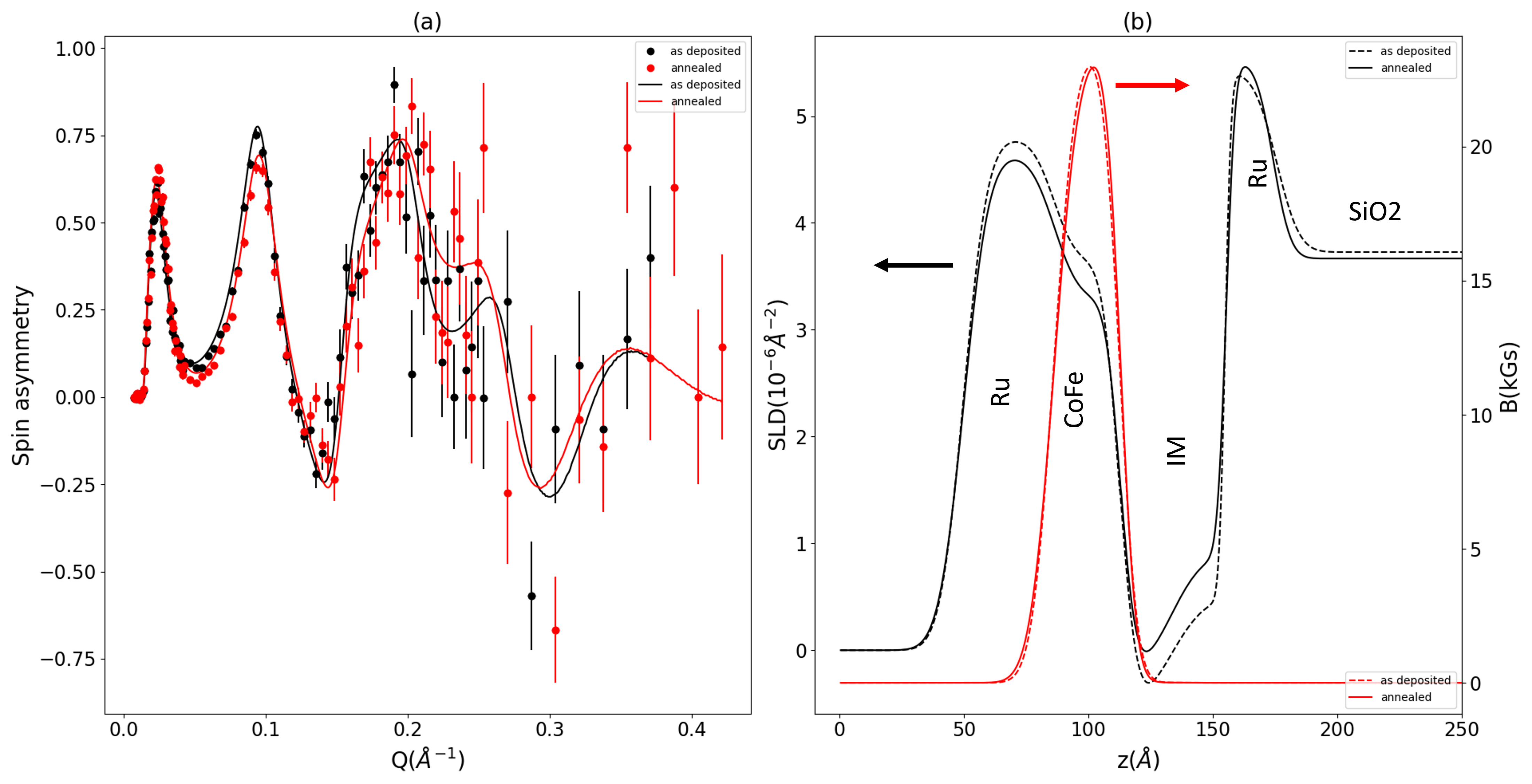}
\caption{(a) Experimental (dots) and model (solid lines) spin asymmetries of the pristine (CF0, black) and 300°C annealed (CF2, red) samples. (b) Corresponding nuclear (black) and magnetization (red) depth profiles before (dashed lines) and after (solid lines) annealing.}
\label{fig3}
\end{figure}

Fig.\ref{fig3}a shows the spin asymmetries before and after annealing. The strongest changes occur in the region of the first maximum and consist in a decrease in the oscillation amplitude after annealing. In the model we can describe it by following changes (Fig.\ref{fig3}b): (a) overall increase of SLD of the IM layer and corresponding decrease of SLD of upper CoFe and Ru layer. Magnetic depth profile remains practically intact upon annealing.

\subsection{Samples with Co}
\begin{figure}[ht]
\centering
\includegraphics[width=0.8\textwidth]{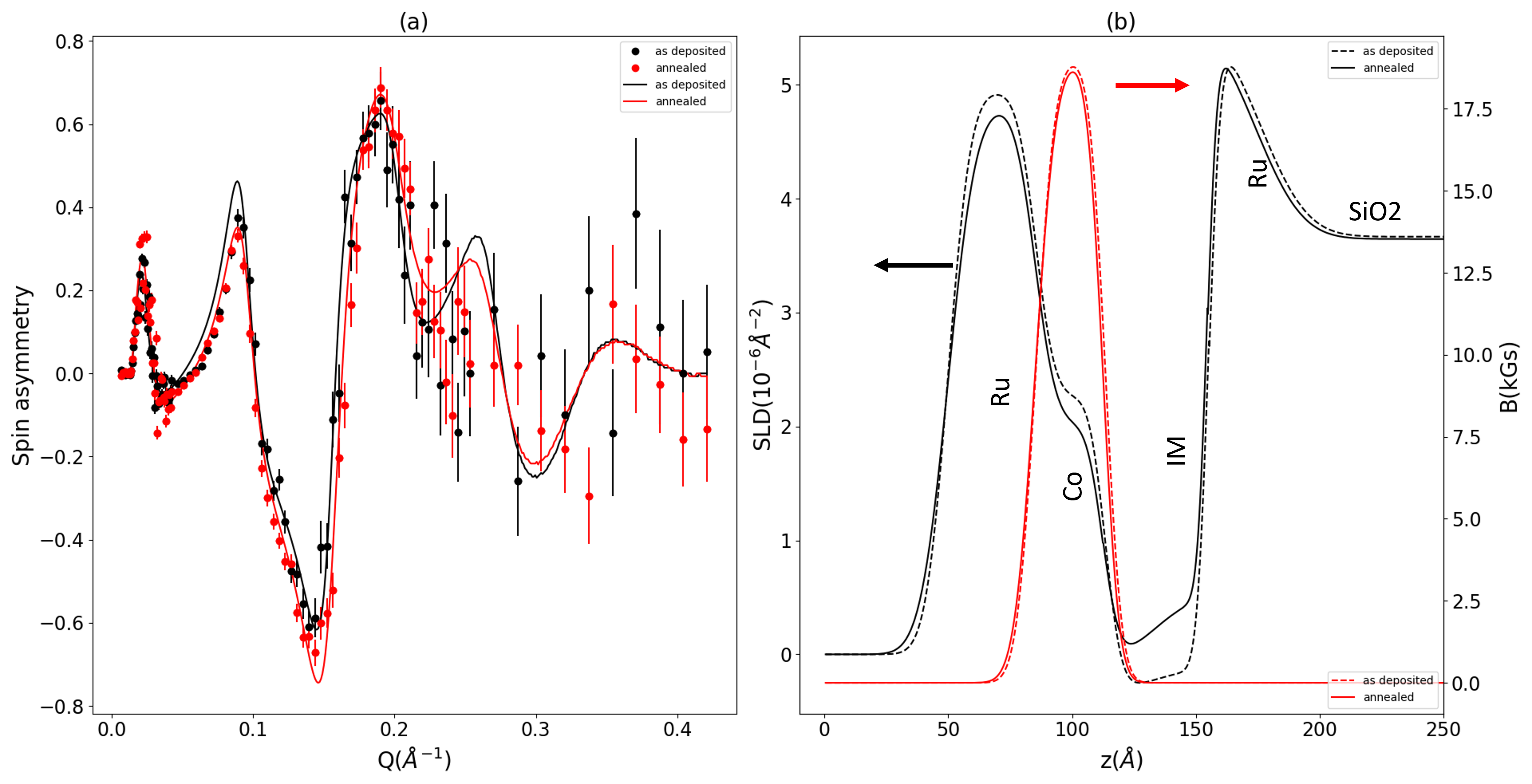}
\caption{(a) Experimental (dots) and model (solid lines) spin asymmetries of the pristine (Co0, black) and 300°C annealed (Co2, red) samples. (b) Corresponding nuclear (black) and magnetization (red) depth profiles before (dashed lines) and after (solid lines) annealing.}

\label{fig4}
\end{figure}

Data for samples with Co layers are shown in Fig.\ref{fig4}. As can be seen from the figure, annealing of systems with cobalt layers is qualitatively similar to the behavior of systems with CoFe layers – decrease (increase) of SLDs of Ru/Co (IrMn) bilayer and small suppression of magnetization of Co layer. 

\subsection{Samples with Fe}

\begin{figure}[ht]
\centering
\includegraphics[width=0.8\textwidth]{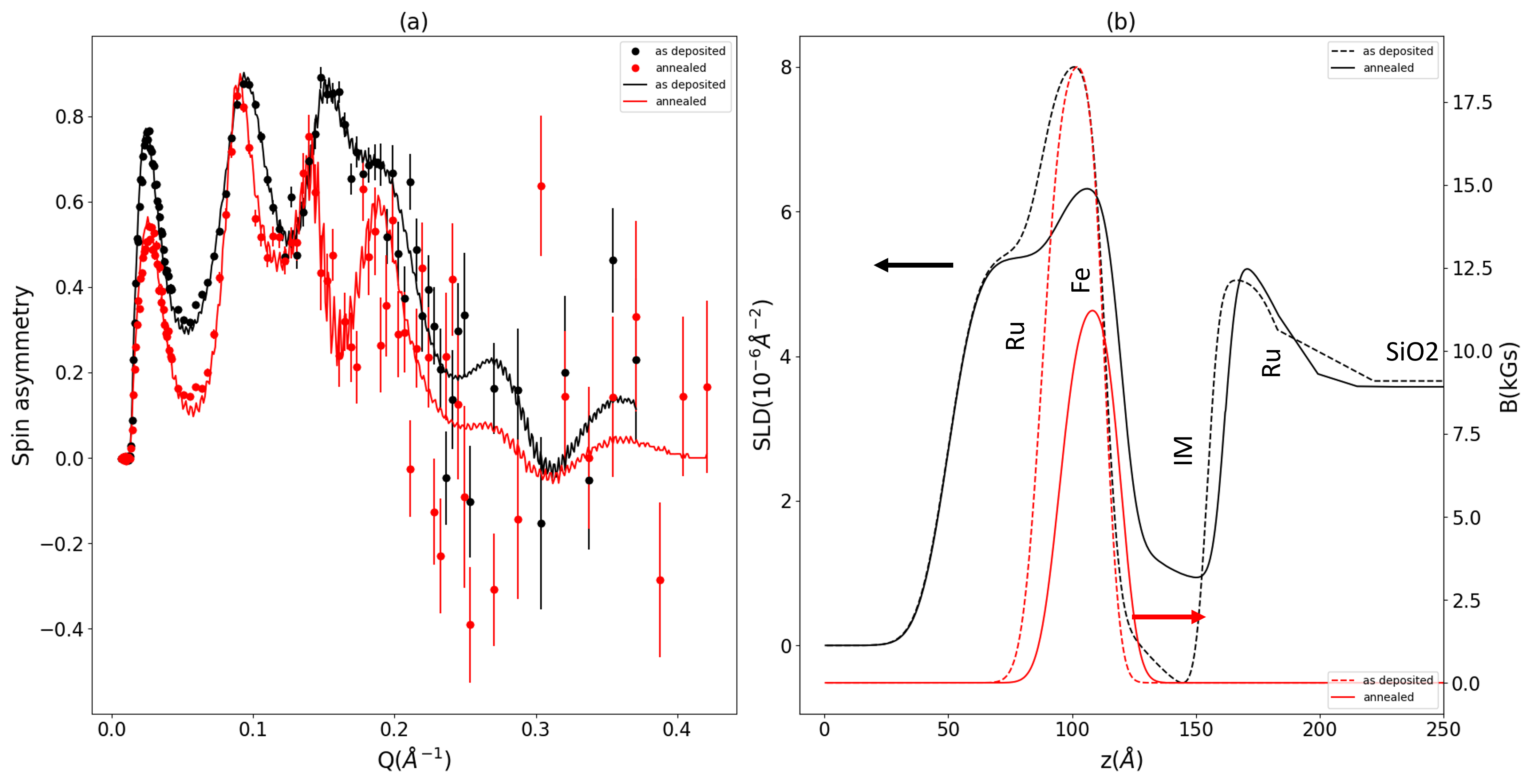}
\caption{(a) Experimental (dots) and model (solid lines) spin asymmetries of the pristine (Fe0, black) and 300°C annealed (Fe2, red) samples. (b) Corresponding nuclear (black) and magnetization (red) depth profiles before (dashed lines) and after (solid lines) annealing.}
\label{fig5}
\end{figure}

A completely different annealing pattern is observed for samples with Fe layer (Fig. \ref{fig5}). A strong suppression of the amplitude of the SA oscillations is observed in the experiment (Fig.\ref{fig5}a). The corresponding model calculations (Fig.\ref{fig5}b) show that this behavior can be explained by a decrease in the magnetization of the iron layer from 19kG to 11kG, as well as a change in the nuclear profile - a decrease in the SLD of the iron layer and an increase in the AFM of the layer. This behavior indicates a strong interdiffusion of iron and manganese atoms. The shift of the interface may indicate a predominant diffusion of iron into the AFM layer.

\section{XMCD results}

\begin{figure}[ht]
\centering
\includegraphics[width=1.0\textwidth]{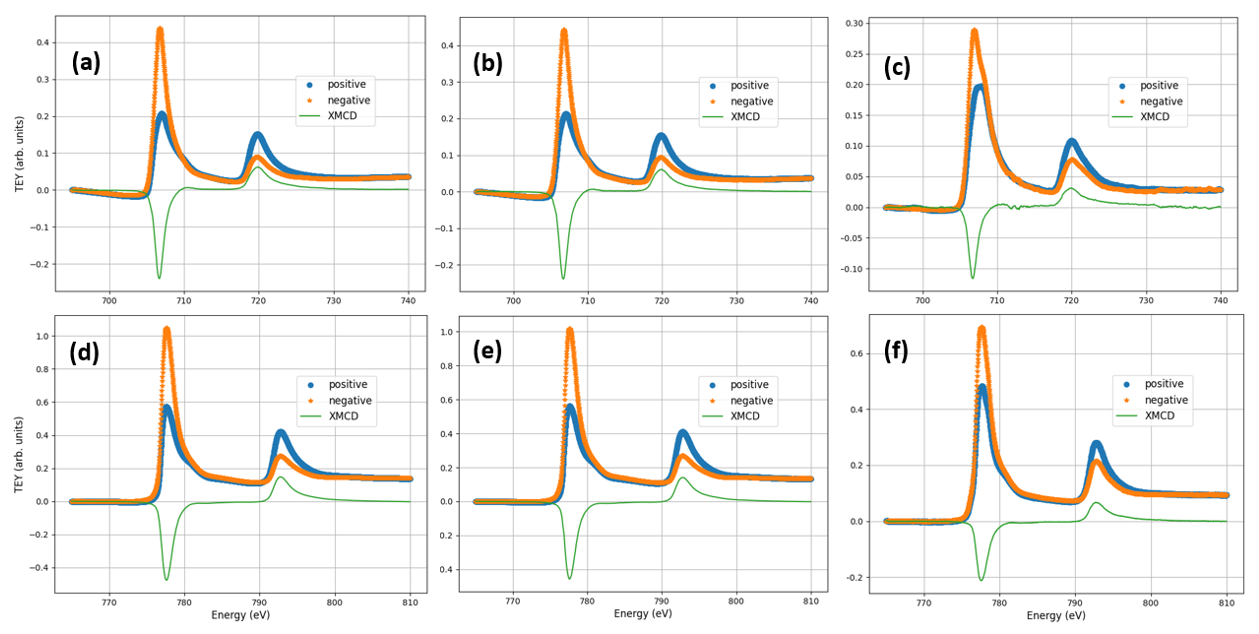}
\caption{XAS spectra measured with positive (blue circles) and negative (orange stars) circular polarization and corresponding XMCD signals (green line) measured across Fe L2/3 edges (top row) and Co L2/L3 edges (bottom row) for the IM/Co$_{70}$Fe$_{30}$ samples as deposited (a, d), annealed at 250°C (b, e), and annealed at 300°C (c, f). }
\label{fig6}
\end{figure}

Fig.\ref{fig6} shows the XAS spectra taken on the samples with the Co$_{70}$Fe$_{30}$ ferromagnetic layer, measured across Fe (top row) and Co (bottom row) L2/3 absorption edges. The left column shows the spectra of the as deposited sample, middle column – data for sample annealed at 250°C, and right column – sample annealed at 300°C. For Co the spectra look all alike, indicating that no significant intermixing involving Co occurs. For Fe as deposited and annealed at 250°C spectra are similar as well, while for the sample annealed at 300°C  emergence of a shoulder at the L3 edge can be seen (Fig.\ref{fig8}c), suggesting developing of a multiplet structure due to changing of chemical environment of Fe atoms. 

\begin{figure}[ht]
\centering
\includegraphics[width=1.0\textwidth]{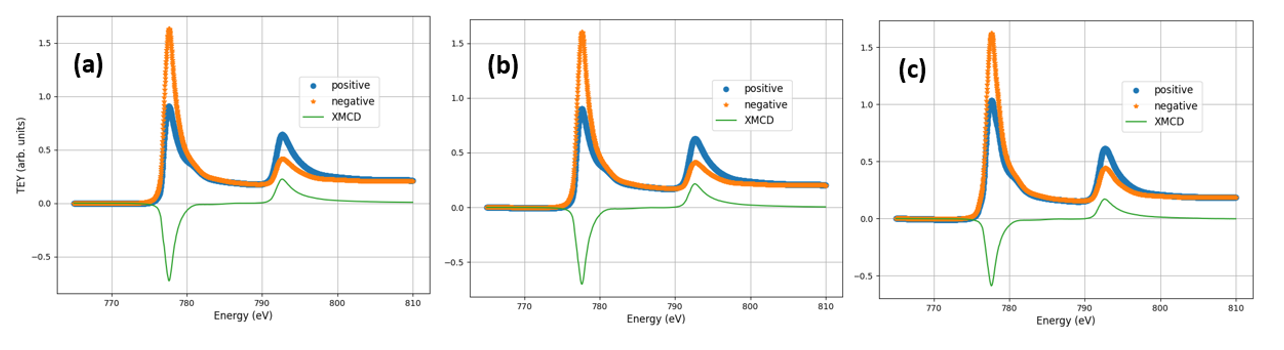}
\caption{XAS spectra measured with positive (blue circles) and negative (orange stars) circular polarization and corresponding XMCD signals (green line) measured across Co L2/3 edges for the IM/Co samples as deposited (a), annealed at 250°C (b), and at 300°C (c).}
\label{fig7}
\end{figure}

\begin{figure}[ht]
\centering
\includegraphics[width=1.0\textwidth]{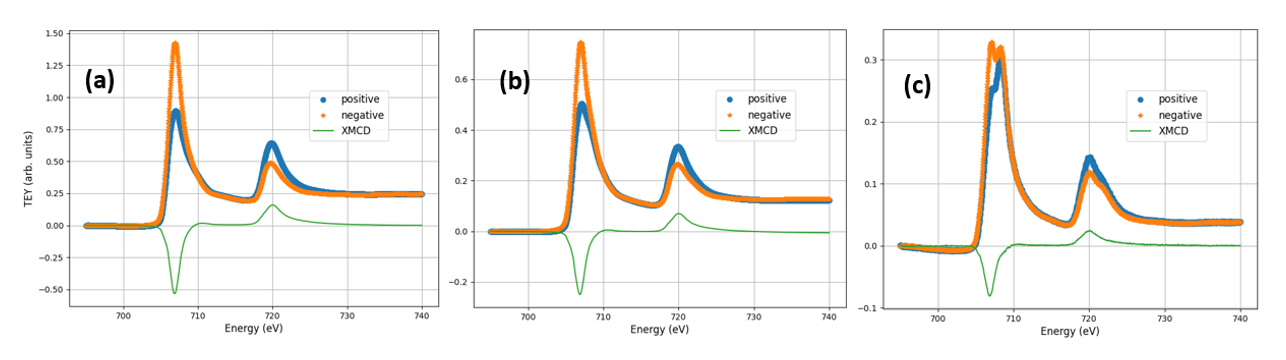}
\caption{XAS spectra measured with positive (blue circles) and negative (orange stars) circular polarization and corresponding XMCD signals (green line) measured across Fe L2/3 edges for the IM/Fe samples as deposited (a), annealed at 250°C (b), and at 300°C (c).}
\label{fig8}
\end{figure}

\begin{figure}[ht]
\centering
\includegraphics[width=1.0\textwidth]{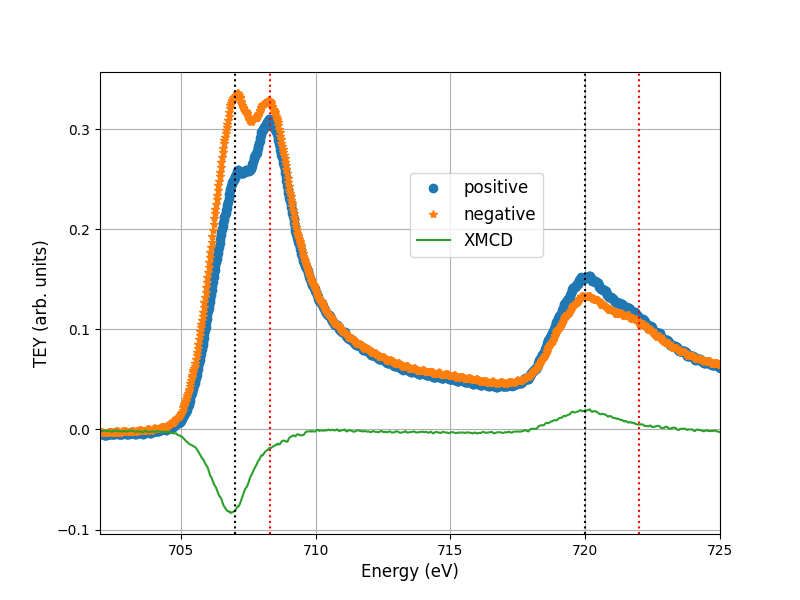}
\caption{Zoomed in XAS and XMCD spectra of the IF/Fe sample annealed at 300°C. Multiplet structure (black and red dotted lines) is developed due to strong interfacial intermixing, forming two distinct environments for Mn: 'old' at  (black dotted lines), existing in other samples, and providing uncompensated magnetic moment (XMCD peaks), and 'new', having zero net magnetic moment (no XMCD peaks).}
\label{fig8a}
\end{figure}

Fig.\ref{fig7} shows the corresponding spectra for the sample containing Co, and no visible change occurs with annealing. Fig.\ref{fig8} contains such spectra for the IrMn/Fe samples. While no significant changes can be observed for the sample annealed at 250°C, for the one annealed at 300°C a well defined multiplet structure develops in XAS spectra (Fig.\ref{fig8}c). Zoomed in spectra for this sample are shown in Fig.\ref{fig8a}. Two distinct environments of Mn are evident, with one (peaks marked with black dotted lines) contributing to the XMCD signal due to uncompensated interfacial Mn moments, coupled to the Fe layer, while the other one (peaks marked with red dotted lines), having no net magnetic moment.   A similar but less pronounced non-magnetic shoulder can be seen developing on the IrMn/CoFe sample annealed at 300°C (Fig.\ref{fig6}c) at 708.3 eV. This is consistent with PNR and magnetization measurements, which showed significant interfacial intermixing and reduction of magnetization for the IrMn/Fe sample annealed at 300°C, as well as with previous works, where formation of interfacial FeMn was observed in IrMn/CoFe systems after etching or annealing at high temperatures \cite{ODonnell2019,Salazar2018}. More precise identification of the observed chemical shift in this sample would require more XANES measurements and multiplet simulations, but it is beyond the scope of this paper.

Total electron yield detection of the XMCD allows for quantitative analysis of spin and orbital moments \cite{Laan_2013}. While the exact values of these moments would require multiplet simulations to determine the shape of the L3 / L2 transition function, the ratio of orbital to spin moments can be found from the XMCD sum rules using the following expression \cite{Thole_XMCD, Chen_SumRules}:   

\begin{equation}
\frac{m_{\mathrm{orb}}}{m_{\mathrm{spin}}} = - \frac{\frac{4}{3} \int_{L_3 + L_2} (\mu_+ - \mu_-) \, dE}{  6 \int_{L_3} (\mu_+ - \mu_-) \, dE - 4 \int_{L_3 + L_2} (\mu_+ - \mu_-) \, dE }
\end{equation}

Here $\mu_+$ and $\mu_-$ denote XAS intensities for positive and circular polarisation respectively, and integration is performed either over the $L_3$ absorption edge, or over both $L_3$ and $L_2$  edges, as shown in the equation. The extracted values of Fe and Co orbital to spin moment  ratios for all samples  are summarised in Table \ref{L2Stab}.

\begin{table}[ht]
\centering
\caption{Orbital to spin moment ratios extracted from XMCD sum rules.}
\label{L2Stab}
\begin{tabular}{|c|c|c|c|}
\hline
    Annealing temperature & No annealing & 250$^\circ$C & 300$^\circ$C \\ \hline
    Fe FM layer  & 0.04 & 0.04 & 0.22 \\ \hline
    Co FM layer  & 0.11 & 0.11 & 0.14 \\ \hline
    CoFe FM layer & 0.12 (Fe), 0.14 (Co) &  0.14 (Fe), 0.14 (Co) &  0.11 (Fe), 0.16 (Co) \\ \hline
    
  \end{tabular}
 \end{table}

No significant changes in the orbital to spin moment ratios can be observed, except from its 5 fold increase in the Fe sample annealed at 300$^\circ$C, consistent with the symmetry reduction due to strong interdiffusion \cite{Laan_2013}.

\section{Discussion}
The PNR data for as-deposited samples reveal a non-uniform SLD depth profile within the IM layer, indicative of a depth-dependent stoichiometry.For the Co0 and CF0 samples, the nature of the profile—more negative towards the surface—suggests a propensity for manganese to migrate upwards.  Utilizing the IM density 8 g/cm$^3$ estimated from X-ray SLDs in Ref.\cite{Fan22}, we can estimate the stoichiometry at the CoFe/IM and IM/Ru interfaces to be IrMn${3.5}$ and IrMn${2}$, respectively. The diffusion of manganese is likely driven by its oxidation affinity, causing it to move towards the surface \cite{Lee01,Lee02}. Conversely, for the Fe0 sample, a more negative SLD is observed at the IM/Ru interface. The annealing of the sample with iron, according to PNR data, leads to significant mixing in the system. The shift of the interface boundary and the increase in the SLD of the AFM layer indicate the predominant penetration of iron into the IM layer, likely due to iron’s tendency to form an FeMn alloy, as evidenced by the split peak in the XMCD (Fig.\ref{fig8a}). Additionally, reduced exchange energy obtained from magnetometric measurements, indicates a strong degradation of the Fe/IM barrier. 

\begin{figure}[ht]
\centering
\includegraphics[width=0.9\textwidth]{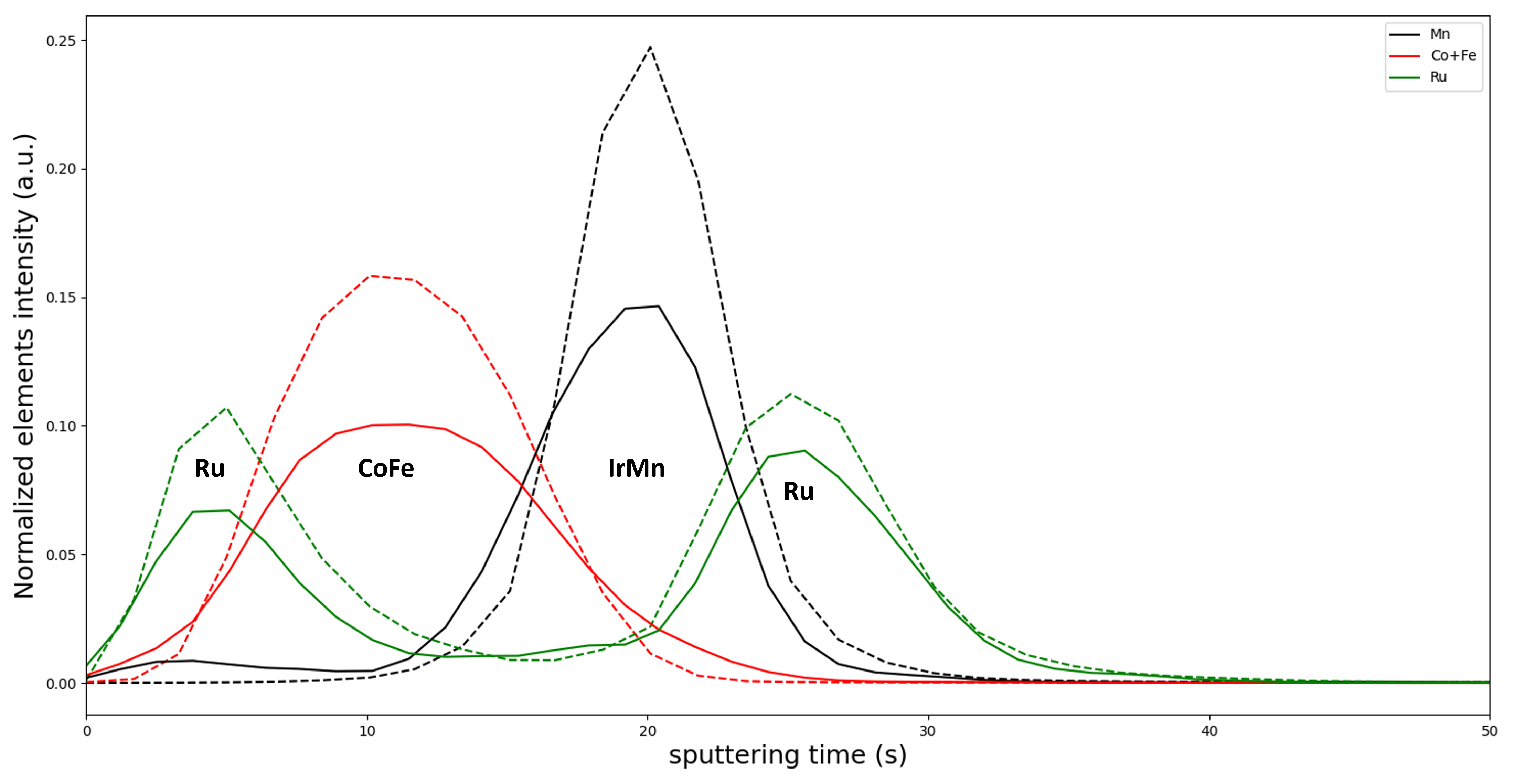}
\caption{SNMS spectra for the not annealed sample (S4, dashed) and annealed (S5, solid line). Each element intensity was normalized on corresponding integrated signal. }
\label{fig9}
\end{figure}

Another picture is observed in Co and CoFe samples. We see an increase of IM layer SLD together with suppression of SLD of the FM and capping layer. The reduction in the ferromagnetic layer’s SLD could theoretically be linked to a decrease in average density. However, this would lead to a diminished magnetization of the ferromagnetic layer, which is not evident in the experiment. An alternative explanation could be the continued migration of manganese towards the surface. To test this hypothesis, additional concentration profile measurements were conducted using Secondary Neutral Mass Spectrometry (SNMS). Figure\ref{fig9} illustrates the normalized intensity distributions for Mn, Co+Fe, and Ru peaks. As can be seen from the figure, annealing indeed leads to observable migration of Mn towards the surface. From decrease of SLDs of Ru, Co and CoFe layers we can estimate Mn concentration in corresponding layers of CF2 and Co2 samples as $\sim$ 5 \%. 

In conclusion, our investigation into the annealing effects on IrMn/FM (FM=Co, Fe, CoFe) structures through PNR, XMCD, magnetometry, and SNMS has revealed nuanced responses to thermal treatment. The Co and CoFe samples exhibit robustness against annealing up to 300°C, maintaining their magnetic integrity with minimal diffusion, except for a notable migration of Mn ($\sim$ 5 \%) towards the surface. On the other hand, the Fe sample demonstrates a pronounced degradation at the Fe/IrMn interface, suggesting the formation of an FeMn alloy, as evidenced by the shift in interface boundary and increased SLD in the AF layer. This is further corroborated by magnetometric data indicating a weakened Fe/IM barrier and XMCD results showing a split peak, indicative of alloying. The comprehensive data, including SNMS profiles, confirm the complex interplay of elemental migration and interaction under annealing conditions, impacting the structural and magnetic properties of these layered systems. 

\section{Acknowledgements}
AV thanks the Swedish Research Council, VR for financial support (Project No. 2017-00651). ILL is acknowledged for providing neutron beam time on Super ADAM (doi:10.5291/ILL-DATA.5-54-285).

%\section*{References}
\printbibliography
\end{document}